\documentclass[%
 reprint,
groupedaddress,
showpacs,preprintnumbers,
 amsmath,amssymb,
 aps,
]{revtex4-1}

\usepackage{graphicx}
\usepackage{dcolumn}
\usepackage{bm}
\usepackage{subfigure}

\begin{document}

\preprint{APS/123-QED}

\title{Multi-configuration Dirac-Hartree-Fock calculations of excitation energies, oscillator strengths and hyperfine structure constants for low-lying levels of Sm I}

\author{Fuyang Zhou}
\author{Yizhi Qu}
\address{College of Material Sciences and Optoelectronic Technology, University of Chinese Academy of Sciences, Beijing 100049, China}
\author{Jiguang Li}\email{li\_jiguang@iapcm.ac.cn}
\author{Jianguo Wang}
\address{Data Center for High Energy Density Physics, Institute of Applied Physics and Computational Mathematics, Beijing 100088, China}

\date{\today}

\begin{abstract}
The multi-configuration Dirac-Hartree-Fock method was employed to calculate the total and excitation energies, oscillator strengths and hyperfine structure constants for low-lying levels of Sm I. In the first-order perturbation approximation, we systematically analyzed correlation effects from each electrons and electron pairs. It was found that the core correlations are of importance for physical quantities concerned. Based on the analysis, the important configuration state wave functions were selected to constitute atomic state wave functions. By using this computational model, our excitation energies, oscillator strengths, and hyperfine structure constants are in better agreement with experimental values than earlier theoretical works.

\end{abstract}

\pacs{31.15.ve,31.15.vj,31.30.Gs}
\maketitle


\section{Introduction}

\begin{figure}[b]
\includegraphics[width=0.5\textwidth]{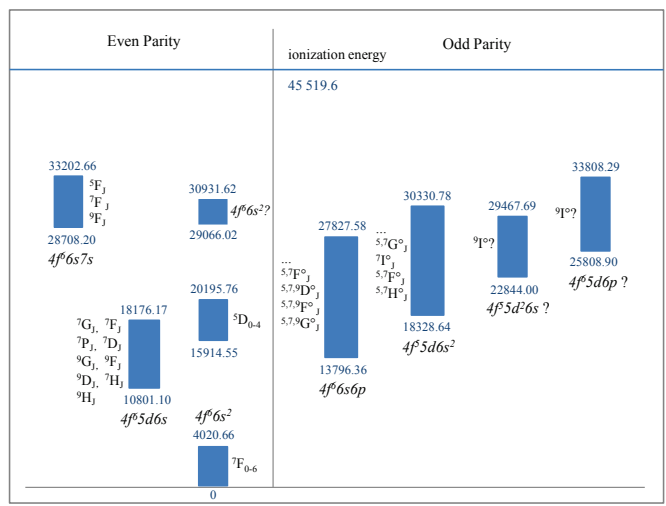}\\
\caption{\label{fig:1}(Color online) Scheme of the odd-parity and even-parity levels (in unit cm$^{-1}$) of the Sm atom.}
\end{figure}

The complicated electronic structure of lanthanide atoms leads to unique physical properties which are of great interest to various applications. For example, the lanthanide luminescence was investigated for biomedical analyses and imaging in view of their enabling easy spectra and time discrimination of the emission bands which span both the visible and near-infrared ranges~\cite{Bunzli2010}. The rich and broad spectra of rare earth elements are also accessible to astronomy studies~\cite{Lawler2009,Sneden2009}, and have many applications in the lighting community~\cite{B406082M}. However, investigation of atomic parameters for rare earth atoms are quite difficult due to the complicated and strong electron correlation effects mainly arising from electrons in the $4f$ open shell \cite{Martin1978,Beck2009,Beck2010}. Among the lanthanide atoms, the energy level structure of samarium is one of the most complex, as shown in Fig~\ref{fig:1}~\cite{NIST_ASD}. The term of its ground configuration is [Xe] $4f^66s^2~^7F$. The open $4f, 5d, 6s$, and $6p$ shells in excited states give rise to the complex structure of energy levels, and large overlap between energy blocks can be found from this figure. In this work, we focus on the transitions from configuration $4f^66s^2$ to $4f^66s6p$, which contain the lowest states for odd and even parities.
%

Because of the strong correlation effects, there are only a few \textit{ab-initio} calculations of atomic properties for Sm I. In 1997, Porsev~\cite{Porsev1997} studied the lifetimes of low-lying odd-parity levels $4f^66s6p~^9G^o_{0-4}$ and $^9F^o_{1,2}$ within the framework of the relativistic configuration interaction (RCI) method, where electron correlations involving the valence orbitals ($4f, 5d, 6s$ and $6p$) were considered. The deviation of their excitation energies from experimental values~\cite{Blagoev1977,Lawler2013} is about 15\%, and the calculated oscillator strengths are not consistent with the experimental values~\cite{Komarovskii1996,Komarovskii1970} either. Latter, based on the multi-configuration Dirac-Hartree-Fock (MCDHF) method, Dilip \textit{et al.}~\cite{Dilip2001} investigated the excitation energies and the hyperfine structure (HFS) constants for low-lying levels, where the active space approach were adopted to include the valence correlations involving $5d, 6s$ and $6p$ electrons. Their excitation energy values are very close to Porsev's.  For HFS, it is interesting that the discrepancy between their results and experimental measurement~\cite{Childs2011} is considerably large, while the results in single-configuration approximation by Cheng and Childs~\cite{Cheng1985} agree with experiment quite well. In their calculations, due to the limited computational capacity at that time, the core correlations have not been taken into account, although were found to be of importance for heavy elements~\cite{zou2002}.

In this work, we explored the effect of correlation from each electron pair on total energies, excitation energies and HFS constants for low-lying levels of Sm within the framework of the MCDHF method~\cite{I.P.Grant2007}.
On the basis of analysis of electron correlations, the important configurations, accounting for main electron correlations in the first-order perturbation approximation, were selected to calculate the different atomic properties. The agreement between present results and experimental values was dramatically improved by including core correlations.

\section{Theory}
In the MCDHF approach, the atomic state wave function (ASF) $\Psi$ is represented as a linear combination of symmetry-adapted configuration state wave functions (CSFs) $\Phi$
\begin{eqnarray}
\Psi \left( {\Gamma \pi JM} \right) = \sum\limits_r {c_\Gamma ^r} \Phi \left( {{\gamma _r} \pi JM} \right),
\end{eqnarray}
where $\pi$, $J$ and $M$ are the parity, total angular momentum and magnetic quantum number, respectively. $\Gamma$ and $\gamma _r$ are the additional quantum numbers to define each ASF or CSF uniquely. 
Configuration mixing coefficients $c_\Gamma ^r$ are obtained through diagonalization of the Dirac-Coulumb Hamiltonian
\begin{eqnarray}
\label{eqn:2}
{H_{DC}} = \sum\limits_{i = 1}^N {[c{\bm \alpha _i} \cdot {\bm p_i} + \left( {{\beta _i} - 1} \right){c^2} + V({r_i})]}  + \sum\limits_{i > j} {\frac{1}{{{r_{ij}}}}} ,
\end{eqnarray}
where the $V(r_i)$ is the monopole part of the electron-nucleus Coulomb interaction, $\bm \alpha _i$ and $\beta_i$ are the Dirac matrices. In the relativistic self-consistent field  procedure, both the radial parts of Dirac orbitals and the expansion coefficients $c_\Gamma ^r$ are optimized~\cite{Fischer1997}.

The Breit interaction in the low frequency approximation
\begin{eqnarray}
{B_{ij}} =  - \frac{1}{{2{r_{ij}}}}\left[ {{\bm \alpha _i} \cdot {\bm \alpha _j} + \frac{{({\bm \alpha _i} \cdot {\bm r_{ij}})({\bm \alpha _j} \cdot {\bm r_{ij}})}}{{r_{ij}^2}}} \right]
\end{eqnarray}
and the QED effects including vacuum polarization and self-energy correction can be included in the relativistic configuration interaction computations\cite{Jonsson2013,Jonsson2007,Parpia1996}.

Once the initial and final state wave functions have been obtained, the radiative transition matrix element can be expressed as
\begin{eqnarray}
\ {M_{if}} = \langle \Psi(i)\| \ {O^{(1)}} \| \Psi(f)\rangle.
\end{eqnarray}
Here $O^{(1)}$ is the electric dipole (E1) interaction. The standard Racah algebra assumes that the orbital sets for the initial- and final-state wave functions are the same \cite{Fano1965}. This restriction can be relaxed by the biorthogonal transformation technique \cite{Olsen1995}. As a result, the transition matrix elements described by independently optimized orbital sets, can also be calculated using Racah algebra.

The hyperfine structure of the atomic energy levels is caused by electromagnetic interactions between the nucleus and  electrons. The magnetic dipole (M1) and electric quadrupole (E2) hyperfine interaction constants $A$ and $B$ 
are given by~\cite{Jonsson1996}
\begin{eqnarray}
\ {{{A}}_J} = \frac{{{\mu _I}}}{I}\frac{1}{{{{[J(J + 1)]}^{1/2}}}}\left\langle {{\Gamma _J}J\left\| {{\bm T^{(1)}}} \right\|{\Gamma _J}J} \right\rangle
\end{eqnarray}
and
\begin{eqnarray}
\ {{{B}}_J} = 2{Q_I}\left[ {\frac{{J(2J - 1)}}{{(J + 1)(2J + 3)}}} \right]\left\langle {{\Gamma _J}J\left\| {{\bm T^{(2)}}} \right\|{\Gamma _J}J} \right\rangle.
\end{eqnarray}
Here, $I$ is the nuclear spin, $\mu _I$ is the nuclear magnetic dipole moment, $Q_I$ is the nuclear quadrupole moment, and $\bm T^{(k)}$ is the electronic tensor operators of rank $k$. The M1 and E2 hyperfine operators $\bm T^{(1)}$ and $\bm T^{(2)}$ are defined as~\cite{Jonsson1996}, in atomic units,

\begin{eqnarray}
{\bm T^{(1)}} = \sum\limits_{j = 1}^N {\bm t^{(1)}(j)} =  - i\alpha \left( {{\bm \alpha _j} \cdot \bm 1 _j{\bm C}^{(1)}(j)} \right)} {r_j^{ - 2}
\end{eqnarray}
and
\begin{eqnarray}
{\bm T^{(2)}} = \sum\limits_{j = 1}^N {\bm t^{(2)}(j)} =  - {\bm C^{(2)}(j)}} {r_j^{ - 3},
\end{eqnarray}
where $\alpha$ is the fine-structure constant, and $\bm C^{(k)}$ is the spherical tensor operator of rank $k$.

In this work, the new-version of the GRASP2K package~\cite{Jonsson2013} was adopted to calculate wave functions and atomic properties, such as excitation energies, oscillator strengths and HFS constants.

\section{\label{Sec3} Electron correlation effects}


In the multi-configuration calculations, one can obtain the indication of the important correlation corrections according to the perturbation theory~\cite{Fischer1997,Li2012}. The zero-order ASF should include the dominant configuration state functions. The first-order correction of ASFs is composed of all CSFs which interact with zero-order ASFs, and thus can be expressed as a linear combination of CSFs that are obtained by single and double (SD) substitutions from occupied orbitals of the reference configuration to virtual orbitals.
The first-order correlations can be further classified into different pair correlations which are defined by all possible substitutions from a certain electron pair~\cite{Fischer1997}.
Based on determination of the contributions from each pair correlation, the important correlation corrections can be selected effectively to investigate atomic quantities.



In this work, we were concerned with the low-lying levels of Sm I, the ground configuration of which is [Xe]$4f^66s^2$, and the lowest odd-parity levels belongs to the [Xe]$4f^66s6p$ configuration. The $4f,~6s,~6p$ orbitals were treated as valence, and the others were core orbitals. Due to the huge CSFs space arising from the open $4f$ shell, the first-order correlation effects (especially for the excited state) could not be completely included within our computational capacity. Therefore we divided them into several subsets from individual electrons or electron pairs, the contributions from which could be evaluated in a series of smaller configuration interaction (CI) calculations.
The analysis of the first-order electron correlations for the ground and excited states were proceeded as follows:

(1)	The occupied orbitals were optimized as spectroscopic in the single-configuration approximation and kept frozen in subsequent calculations. The relaxation effect were accounted for by the independent optimization of the ground and excited states.

(2) The virtual orbitals were generated in an restricted configuration space, in which only some electron-pair correlations were included in the MCDHF approach. For example, in the relativistic self-consistent field procedure, the configurations could be obtained by single and double (SD) substitutions from valence orbitals to the virtual ones. As a result, these virtual orbitals were optimized to accommodate the contributions from only valence correlations.

(3) By applying the orbitals generated above, the differents electron correlation effects could be included in a series of CI calculations to select the important ones.

(4) The Breit and QED corrections were estimated in the single-configuration approximation.

In the evaluations of various atomic properties, the important configuration state wave functions were selected on the basis of analysis of electron correlation effects. Moreover, we could optimize the orbitals to accommodate the contributions from the selected electron correlations in the framework of the MCDHF method.
%

\begin{table}
\caption{\label{tab:table1}%
Correlation energy $\Delta E$ for ground state $4f^66s^2$ $^7F_0$; `SrD' = MCDHF calculation with the configurations were obtained by all single and restricted double (SrD) substitutions to these virtual orbitals (the restriction was described in text); `SD' = CI calculation with configurations were generated by single and unrestricted double (SD) replacement; `Layers' = number of virtual orbitals of a particular symmetry; `NCF' = number of CSFs with J=0,1,2; and `$E(^{7}F_{1}$-$^{7}F_{0})$' = excitation energy in cm$^{-1}$ of $^7F_1-^7F_0$.
}
\begin{ruledtabular}
\begin{tabular}{ccrcc}
 Model & Layers   & NCF      & $\Delta E$(cm$^{-1}$) &$E(^{7}F_{1}$-$^{7}F_{0})$\\
\hline
SrD & ``$1spdfg$'' & 9152      & -27078.48           & 298.16 \\
SrD & ``$2spdfg$'' & 18164     & -28712.10           & 302.47 \\
SrD & ``$3spdfg$'' & 27176     & -28913.40           & 303.75 \\
SD  & ``$3spdfg$'' & 322280    & -32115.94           & 307.40 \\
\multicolumn{4}{c}{Experiment~\cite{Martin1978}}        & 292.58 \\
\end{tabular}
\end{ruledtabular}
\end{table}

\begin{table}
\caption{\label{tab:table2}%
Correlation energy $\Delta E$ for exited state $4f^66s6p$  $^{9}G^o_0$; `NCF' = number of CSFs with J=0,1; `$E(^{9}G^{o}_{0}$-$^{7}F_{0})$' = excitation energy in cm$^{-1}$ of $^9G^{o}_0-^7F_0$.
}
\begin{ruledtabular}
\begin{tabular}{ccrcr}
Model & Layers     & NCF    &$\Delta E$(cm$^{-1}$)&$E(^{9}G_{0}^{o}$-$^{7}F_{0})$\\
\hline
SrD & ``$1spdfg$''   & 51017  & -23559.04         & 10234.69 \\
SrD & ``$2spdfg$''   & 100370 & -25654.01         & 9773.33  \\
SD  & ``$2spdfg$''   & 457452 & -27549.98         & 10394.95 \\
\multicolumn{4}{c}{Experiment~\cite{Martin1978}}& 13796.36 \\
\end{tabular}
\end{ruledtabular}
\end{table}

\subsection{Generation of the Virtual orbitals}
In the present MCDHF approach, to reduce complexity of self-consistent field calculations the virtual orbitals were added layer by layer. The configurations were obtained by single and restricted double (SrD) substitutions from valence orbitals, in which the two occupied orbitals must be replaced by two same virtual orbitals, i.e. only the double substitutions from $4f,6s,6p$ to virtual orbitals $nl^2$ were permitted.
The one-electron energy values of virtual orbitals do not have physical meaning, the properties of virtual orbitals depend on the correlation effects they describes \cite{Bieron2009}. 
In this paper the virtual orbitals were enclosed in quotation marks to avoid confusion from occupied orbitals, and listed by angular symmetry and quantity. For example, ``$2spd1f$'' means two of each of the ``$s$'', ``$p$'', ``$d$'' symmetries and one of ``$f$'' symmetry.
%

In this section, three virtual layers for levels with even parity and two virtual layers for ones with odd parity were generated within the framework of MCDHF method. In order to check the validity of this restriction on double substitutions, contributions from the reduced configurations were added in the CI calculation, where the configurations were obtained by single and unrestricted double (SD) substitutions from valence orbitals to all virtual orbitals generated above.

Table \ref{tab:table1} and \ref{tab:table2} present the correlation energies for the ground state $4f^66s^2$ $^7F_0$ and the excited state $4f^66s6p$ $^9G^o_0$, as well as the excitation energies of $4f^66s^2~^7F_1$ and $4f^66s6p~^9G^o_0$ states. The correlation energy are defined as the total energy difference between the multi-configuration and single-configuration values.
As can be seen from Table \ref{tab:table1}, the calculations with one virtual layer could capture most of the valence correlation effects for the ground state.
In addition, the SrD model can include the valence correlation effects effectively with a smaller number of CSFs. For example, the the difference in $\Delta E$ between the SrD and SD model is 3196 cm$^{-1}$ for the ground state and 3.74 cm$^{-1}$ in the fine structure of the $^7F_1$ level, while the NCF was reduced to 27176 from 322280.

Similar results for the excited state were given in Table \ref{tab:table2}, in which the total and excitation energies were not significantly improved by the expansion of configuration space. However, the excitation energies E($^{9}G_{0}^{o}$-$^{7}F_{0}$) differ from the experimental value by more than 3000 cm$^{-1}$. Therefore, the contributions from the electron correlations which were not included in the generation of virtual orbitals should be evaluated.

\subsection{Contributions from different electron correlations}

In order to include the significant correlations with respect to computational capacity, we divided the one- and two-body electron correlations into several subsets. The CI approach was used to include different correlation effects within the virtual orbital set ``$1spdfg$'', in which orbitals generated above were applied and kept frozen. 
For example, the valence correlation of $4f^66s6p$ could be divided into $4f$, $6s$, $6p$, $4f^2$, $4f6s$, $4f6p$ and $6s6p$ electron correlations. The $4f$ electron correlation effect was expressed as a linear combination of CSFs that are obtained by single replacement $4f$  $\rightarrow$ ``$v$'' (virtual orbital), and the CSFs for the $4f6s$ electron pair correlation were obtained by unrestricted double replacement $4f6s$ $\rightarrow$ ``$vv'$''. Based on these CI calculations, the contributions from each correlation subset to the total energy were evaluated by the absolute value of correlation energies $\Delta E$.

\begin{figure}[!h]
\includegraphics[width=0.53\textwidth]{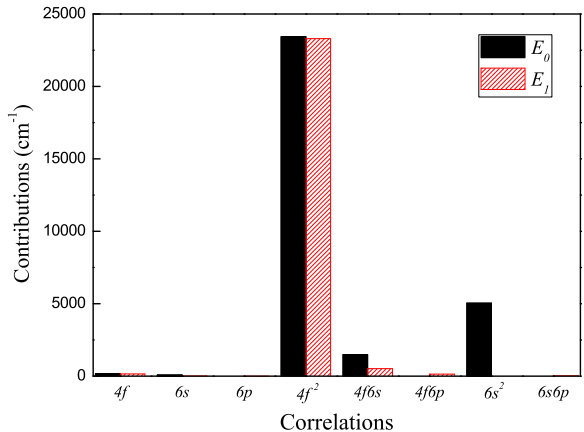}\\
\caption{\label{fig:2}(Color online) Contributions from the different valence correlation effects to the total energies for ground state $4f^66s^2~^7F_0$ ($E_0$) and exited state $4f^66s6p~^9G^{o}_0$ ($E_1$). The contributions were evaluated by the absolute value of correlation energies $\Delta E$.}
\end{figure}


\subsubsection{valence correlations}

The contributions from the different valence correlations to the total energy $E_0$ of the ground state $4f^66s^2~^{7}F_{0}$ and the excited state $4f^66s6p~^{9}G_{0}^{o}$ ($E_1$) were presented in Fig.~\ref{fig:2}.
For the ground state total energy $E_0$, the individual electron correlations are $4f$, $6s$, $4f^2$, $4f6s$ and $6s^2$. It was found that the one-body correlation effects from valence orbitals are negligible. The most important valence correlation is from the $4f^2$ electron pair, while the contributions from the $6s^2$ and $4f6s$ pair correlations to the correlation energies are much smaller than it.

The contributions of the specific correlation effects for the excited state are similar to the ground state. The one-body correlations also have a negligible effects on the total energy $E_1$. The most important valence correlation is from $4f^2$ electron pair, and the $4f6s$ and $4f6p$ correlations contribute little to the total energy. This result indicates that the electrostatic interaction between the deep-lying $4f$ electron and outer $6s$ and $6p$ electrons is very weak.

Although essentially important for the total energies, it was found that the correlation between $4f^2$ electrons have only a small influence to the excitation energy E($^{9}G_{0}^{o}$-$^{7}F_{0}$). The change in excitation energy E($^{9}G_{0}^{o}$-$^{7}F_{0}$) mostly comes from the difference of electron correlation effects involving the external $6s6p$ and $6s^2$ electrons. This could also be due to the weak interaction between the $4f$ electron and outer electrons. 

The excitation energy E($^{9}G_{0}^{o}$-$^{7}F_{0}$) calculated with these VV correlations is 10394.95 cm$^{-1}$ (see Table \ref{tab:table2}), compared with the experimental value 13796.36 cm$^{-1}$. Apart from the valence correlation effects, there are other types of correlation effects involving the core shells.  The difference of 3000 cm$^{-1}$ should come from the core-valence (CV) and core-core (CC) correlation effects.

\begin{figure}[!h]
\includegraphics[width=0.55\textwidth]{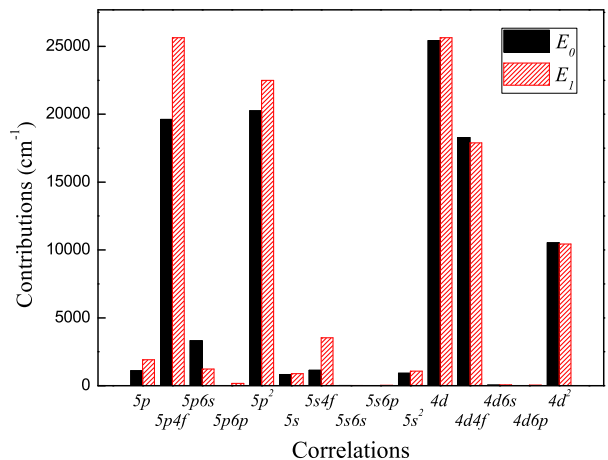}\\
\caption{\label{fig3}(Color online) Contributions from the different electron correlation effects involving core shells $4d5s5p$ to total energies for ground state ($E_0$) and exited state ($E_1$).}
\end{figure}

\begin{figure*}
\includegraphics[width=1\textwidth]{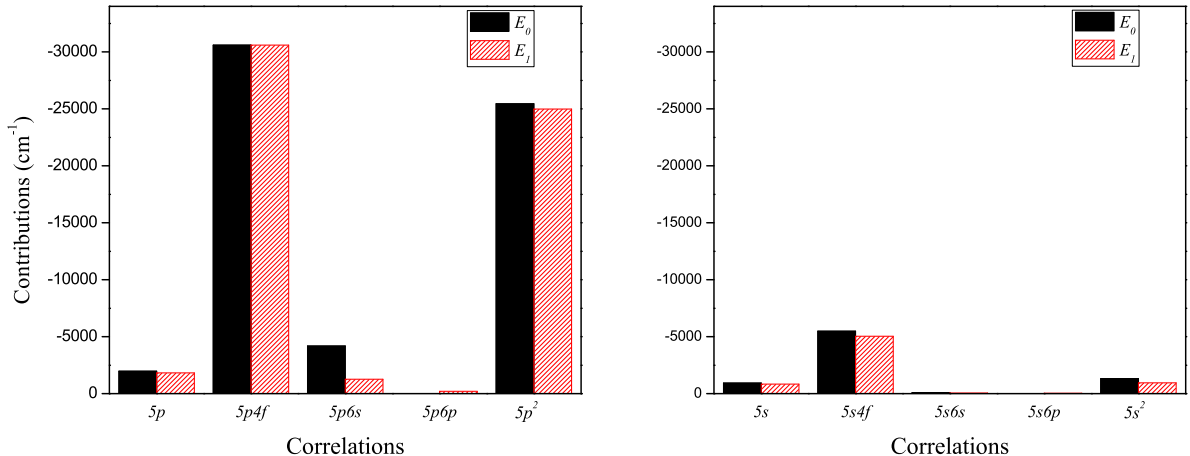}\\
\caption{\label{fig5}(Color online) Contributions from the different correlation effects involving $5p$ and $5s$ electrons to total energies for ground state ($E_0$) and exited state ($E_1$). In order to obtain the accurate contributions for the analysis on excitation energy, the virtual orbitals are re-optimized to accommodate the correlations involving $5p$ and $5s$ electrons, respectively. See text for further details.}
\end{figure*}

\subsubsection{correlations involving core shells}

In Fig \ref{fig3} we presented the contributions from the different correlations involving core shells $4d5s5p$ to the total energy $E_0$ for the ground state $4f^66s^2~^{7}F_{0}$ and the exited state $4f^66s6p~^{9}G_{0}^{o}$ ($E_1$).  The contributions from one-body correlations are quite small except for $4d$ electron. 
In addition, the CV correlations effects between core and $4f$ electrons, especially for the $5p4f$ and $4d4f$ electron pair correlations, are very important to the total energy $E_0$. For CC correlations, only the $nl^2$ pair correlation effects were illustrated in the figure because they are more important than correlations between electrons in different orbitals, e.g., the contributions from $5p^2$ and $4d^2$ pair correlations exceeded 10000 cm$^{-1}$. These correlations involving $4d$ and $5p$ shells strongly influence the atomic state wave functions. The one- and two-body correlation effects involving $3spd4sp$ shells were also taken into account for the ground state, but their contributions are much smaller than $4d5s5p$ shells.

For the exited state, due to the huge CSFs space arising from the reference configuration $4f^66s6p$, only CV and the most important part of CC correlations (i.e. the $nl^2$ pair correlations) were considered.
The effects of electron correlations involving $4d5sp$ shells to the total energy are similar to the ground state $^{7}F_{0}$. The $5p4f$, $4d4f$, $5p^2$ and $4f^2$ electron pair correlation effects are most important, and the contributions from the correlations between the core and the $6p$ electrons are even smaller than $6s$ pair correlations.

As can be seen from Fig.~\ref{fig3}, the correlations involving $5p$ and $5s$ electrons have a significant influence on the excitation energy E($^{9}G_{0}^{o}$-$^{7}F_{0}$). However, it seems that these core correlations would decrease the E($^{9}G_{0}^{o}$-$^{7}F_{0}$), while the result without core correlations lowers the experimental value by about 3000 cm$^{-1}$.
This could be due to that the convergence in the present CI calculations was slowed down by the fact that the virtual orbitals were optimized to accommodate the contributions from only valence correlations.


In order to obtain reasonable contributions from the correlation effects involving $5p$ and $5s$ electrons to the excitation energy, we have re-optimized the virtual orbitals in MCDHF calculations with the inclusion of electron correlations involving core orbital $5s$ and $5p$, respectively.
In Fig.~\ref{fig5} we presented the contributions from these electron correlation effects to the total energy for the ground state ($E_0$) and the exited state ($E_1$) with the new set of virtual orbitals. The contributions to the total energy are similar to those using the previous orbital set, that is, the $5p4f$ and $5p^2$ electron correlation effects are most significant. However, the $5p4f$ and $5p^2$ pair correlations have a negligible influence on the excitation energy. 
Additionally, the correlations between $5p$ and external $6s$ and $6p$ electrons are most important to the excitation energy. For the $5s$ electron, the contribution from the $5s4f$ correlation to the excitation energy are also much small, and the correlations between $5s$ and outer $6s6p$ electrons have a negligible effect on both total and excitation energies.

These individual contributions indicate that the core correlations are very important for total energies. For the excitation energy E($^{9}G_{0}^{o}$-$^{7}F_{0}$), the CV correlation effects from $5p$ electron were found to be significant and tend to improve the result without core correlation.

\subsection{Breit and QED corrections}

\begin{table}
\caption{\label{tab:tabley}%
Breit and QED effect on the excitation energies of $4f^{6}6s6p$ $^{9}G_{0,1}^{o}$ and $^{9}F_{1}^{o}$ states of Sm in cm$^{-1}$; `DHF' = the uncorrelated Dirac-Hartree-Fock calculation, `VV' = multi-configurations calculation with valence-valence correlations.
}
\begin{ruledtabular}
\begin{tabular}{crrr}
Model & $^{9}G_{0}^{o}$-$^{7}F_{0}$ & $^{9}G_{1}^{o}$-$^{7}F_{0}$ &$^{9}F_{1}^{o}$-$^{7}F_{0}$\\
\hline
DHF                           & 6694                        & 6899     &7944     \\
VV                            & 10394.95                    & 10601.49 &11624.21 \\
breit correction              & -67                         & -89      &-77      \\
breit \verb+&+ QED correction & -114                        & -136     &-124     \\
Experiment~\cite{Martin1978}  & 13796.36                    & 13999.50 &14863.85 \\
\end{tabular}
\end{ruledtabular}
\end{table}

In Table \ref{tab:tabley} we displayed the excitation energies of $4f^{6}6s6p$ $^{9}G_{0,1}^{o}$ and $^{9}F_{1}^{o}$ states, as well as the correction of Breit interaction and QED effects in the single-configuration approximation. It was found that these high-order corrections have a negligible effect on excitation energies. The remaining discrepancy between calculations and experimental values is mostly attributed to the electron correlations involving core shells which are not included.

\section{Calculation of atomic properties}
\subsection{Excitation Energies and Oscillator Strengths}
\subsubsection{ \label{sec:cm}Calculational model}

\begin{table}[b]
\caption{\label{tab:table8}%
The number of CSFs as the function of the virtual orbital set in the MCDHF calculations with this correlation model(described in Sec. \ref{sec:cm}); ``n$spdf$'' = virtual orbital set; $J^P$ are the total angular momentum ($J$) and parity ($P$) of an atomic state.
}
\begin{ruledtabular}
\begin{tabular}{ccccc}
Model                & $J^P=1^o$ & $J^P=0^e$ & $J^P=1^e$ & $J^P=2^e$  \\
\hline
Single configuration & 252       & 14        & 19        & 37         \\
``$1spdf$''              & 113231    & 2280      & 10440     & 23220      \\
``$2spd1f$''             & 291689    & 5087      & 26120     & 60421      \\
\end{tabular}
\end{ruledtabular}
\end{table}

As mentioned above, the first-order correlations couldn't be adequately included in the calculations for Sm with respect to the open $4f$ shell that result in a huge CSFs space.
In order to carry out an accurate calculation of the excitation energies and transition probabilities, the important specific correlations should be selected to form the ground and excited atomic state wave functions, based on analysis of electron correlation effects. Also, the orbitals need to be optimized to accommodate the contributions from the selected electron correlations in the framework of the MCDHF method.

As discussed in Sec.~\ref{Sec3}, the valence correlations involving outer $6s$ and $6p$ electrons should be included in this correlation model, which significantly influence the excitation energy. Apart from the valence correlation, the CV correlations was found to be very important as well. 
Therefore, in our MCDHF calculations, the $4f6s$, $4f6p$, $6s^{2}$, $6s6p$, $5p6s$ and $5p6p$ electron pair correlations should be included.

For the transitions from $4f^{6}6s6p$ $^{9}G_{1}^{o}$ and $^{9}F_{1}^{o}$ to $4f^{6}6s^{2}$ $^{7}F_{0,1,2}$, the levels belonging to the lower and upper configurations were optimized in two separate MCDHF calculations. At the starting point, the occupied orbitals were obtained in the single-configuration approximation. Then we extended our calculations to include the selected correlations, which further included the important core correlations compared with the calculations in Table~\ref{tab:table2}. The CSFs were obtained by single and double substitutions from the selected electron pair to vitual orbitals.
Although a larger virtual orbital set is more conducive to including the selected correlations, only the ``$2spd1f$'' virtual orbitals were generated in our calculation due to the large size of CSFs.
The number of CSFs within this computational model was given in Table~\ref{tab:table8}.
The configuration space is considerably smaller than the one generated by the conventional active space approach. For example, the number of CSFs with this correlation model for the $^{9}G_{1}^{o}$ and $^9F^{o}_1$ state is 291689, compared to 20701402 CSFs obtained by SD substitutions from orbitals $4df5sp6sp$ to the same orbital set.


\subsubsection{Results}

\begin{table*}
\caption{\label{tab:table9}%
Excitation energies $E$ (in cm$^{-1}$) and oscillator strengths $f$ ($10^{-4}$) for E1 transitions from odd-parity $^{9}G_{1}^{o}$ and $^{9}F_{1}^{o}$ states to even-parity $^{7}F_{0-2}$ states; `B' = Babushkin gauge; `C' = Coulomb gauge; ``n$spdf$'' = virtual orbital set of the MCDHF calculation with the correlation model described in Sec.~\ref{sec:cm}.
}
\begin{ruledtabular}
\begin{tabular}{cllllllllllll}
&\multicolumn{4}{c}{$^{9}G_{1}^{o}$ - $^{7}F_{0}$}&\multicolumn{4}{c}{$^{9}G_{1}^{o}$ - $^{7}F_{1}$}&\multicolumn{4}{c}{$^{9}G_{1}^{o}$ - $^{7}F_{2}$}\\
Model& $E$ &$f_B$&$f_C$&$f_B$/$f_C$&$\Delta E$&$f_B$&$f_C$&$f_B$/$f_C$& $E$ &$f_B$&$f_C$&$f_B$/$f_C$\\
\hline
Single configuration & 6899& 1.66  & 0.07 & 42   & 6621& 0.01        & 0.02  & 0.45      &6107 & 0.60 & 0.01 & 63   \\
``$1spdf$''              &13087& 16.35 & 61.3 & 0.27 &12819& $10^{-6}$   & 0.03  &3*$10^{-5}$&12305& 7.56 & 33.4 & 0.23 \\
``$2spd1f$''             &14265& 11.05 & 22.7 & 0.49 &14001& 0.006 & 0.03  & 0.24            &13488& 4.01 & 9.46 & 0.42 \\

S. G. Porsev~\cite{Porsev1997} &11533&  6.9\footnotemark[1] &      && 11248&0.1\footnotemark[1]&&&10723&2.8\footnotemark[1]&&\\
Experiment~\cite{Blagoev1977,Komarovskii1970}&13999.50&\multicolumn{2}{c}{12.5}&&13706.92&&&&13187.58&\multicolumn{2}{c}{5.7}&\\
\hline \\
&\multicolumn{4}{c}{$^{9}F_{1}^{o}$ - $^{7}F_{0}$}&\multicolumn{4}{c}{$^{9}F_{1}^{o}$ - $^{7}F_{1}$}&\multicolumn{4}{c}{$^{9}F_{1}^{o}$ - $^{7}F_{2}$}\\
Model&$\Delta E$&$f_B$&$f_C$&$f_B$/$f_C$& $E$ &$f_B$&$f_C$&$f_B$/$f_C$&$\Delta E$&$f_B$&$f_C$&$f_B$/$f_C$\\
\hline
Single configuration &7944 & 4.07 & 1.34  & 3.03 &7665 & 5.70 & 3.42  & 1.7  &7151& 1.47 & 1.98 & 0.74 \\
``$1spdf$''              &14127& 46.4 & 134.3 & 0.35 &13859& 42.2 & 118.9 & 0.35 &13345& 4.66 & 10.7 & 0.44 \\
``$2spd1f$''             &15291& 27.8 & 47.6  & 0.58 &15026& 34.2 & 57.7  & 0.59 &14514& 6.54 & 10.0 & 0.65 \\

S. G. Porsev~\cite{Porsev1997}&12674&19.6\footnotemark[1]&&&12389&30\footnotemark[1]&&&11864&9.0\footnotemark[1]&&\\
Experiment~\cite{Blagoev1977,Komarovskii1970}&14863.85&\multicolumn{2}{c}{28.2}&&14571.21&\multicolumn{2}{c}{31.7}&&14051.93&&&\\
\end{tabular}
\end{ruledtabular}
\footnotetext[1]{$f=f_{B}E_{exp}/E_{th}$.}
\end{table*}

\begin{figure}[b]
\includegraphics[width=0.5\textwidth]{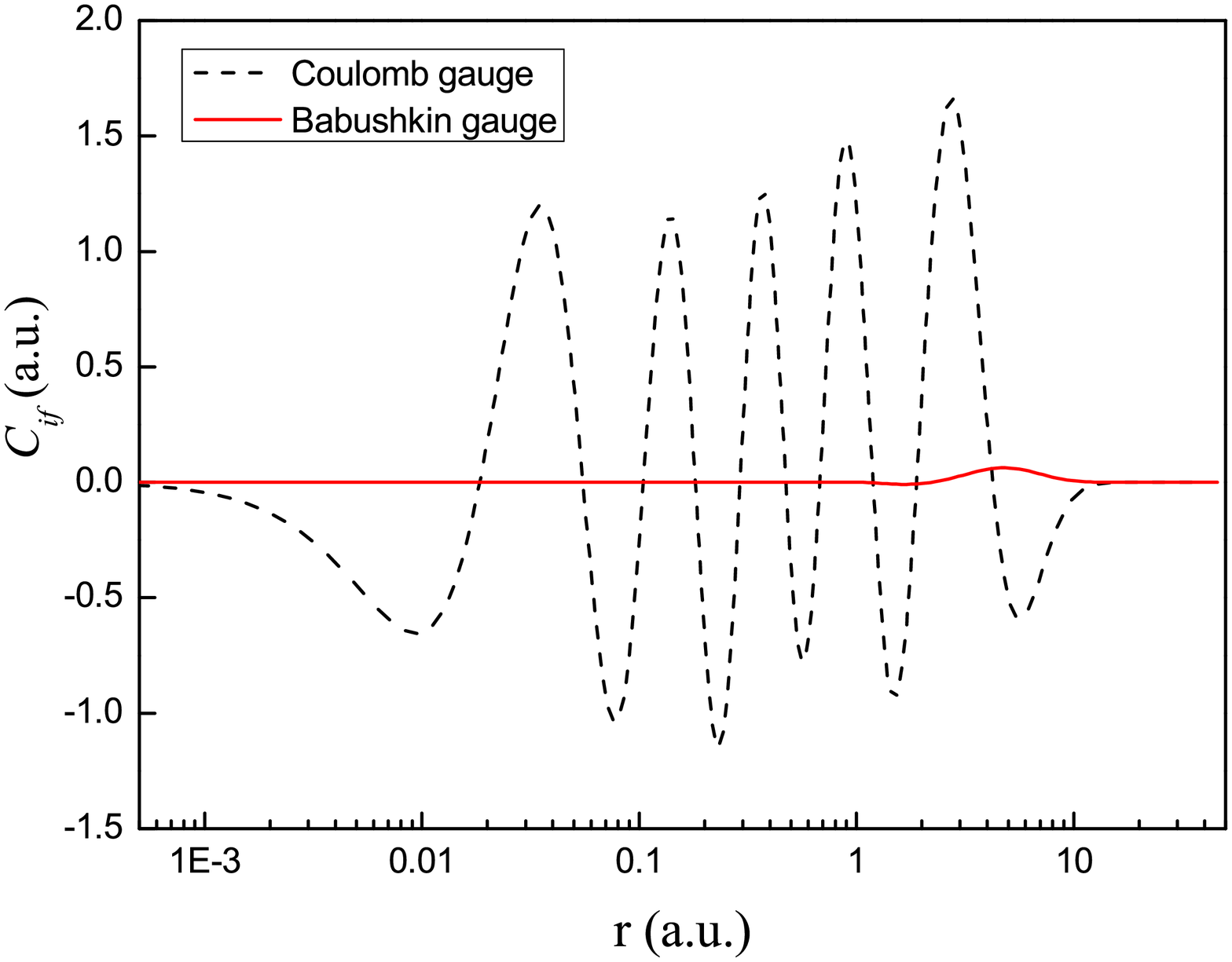}\\
\caption{\label{gauge}(Color online) Radial distribution of the transition matrix element $\langle^7F_0||O^{(1)}||^9G^{o}_1\rangle$ in Coulomb and Babushkin gauges.}
\end{figure}

The transition energies and oscillator strengths of transitions from $4f^{6}6s6p$ $^{9}G_{1}^{o}$ and $^{9}F_{1}^{o}$ to $4f^{6}6s^{2}$ $^{7}F_{0,1,2}$ states for different configuration models were presented in Table~\ref{tab:table9}, and compared with other theoretical and experimental data. The large discrepancy between single-configuration results and experimental values indicates that the strong electron correlation effects exits in Sm I. Due to the computational limitation at that time, only part of the VV correlation could be included in Porsev's CI calculation~\cite{Porsev1997}, and their results of transition energies are about 2000 cm$^{-1}$ lower than experimental data. The present multi-configuration approach gave a much better results. For example, the excitation energy of $4f^66s6p~^9G^{o}_1$ was improved to 14265 cm$^{-1}$, only 266 cm$^{-1}$ higher than the experimental value.
This result indicates that this correlation model could account for the major difference of the electron correlation effects between ground and exited states.

The oscillator strengths in Babushkin (length) and Coulomb (velocity) gauges  ($f_B,f_C$) for transitions from $4f^{6}6s6p$ $^{9}G_{1}^{o}$ and $^{9}F_{1}^{o}$ to $4f^{6}6s^{2}$ $^{7}F_{0,1,2}$ states were also given in Table~\ref{tab:table9}. It can be seen that the electron correlation effects on the oscillator strengths are remarkable. 
Compared with Porsevs's results, the oscillator strengths in Babushkin gauge agree well with experimental data, and the agreement were improved with increase of configuration space. For the transitions $^9F^{o}_1$~-~$^7F_{0,1}$, which have the largest transition probabilities, the deviations are less than 10\% from the experimental values. However, this result couldn't provide the exact transition probabilities for the lines without experimental value, since the values are too small compared with the errors of other lines, especially for the transition $^9G^{o}_1$~-~$^7F_{1}$.  

Meanwhile, we noted that the inconsistency in the oscillator strengths between different gauges is very large. The oscillator strengths in Coulomb gauge is much larger than those in Babushkin gauge. The gauge difference are ascribed to the fact that the E1 transition amplitudes in Babushkin and Coulomb gauges are sensitive to different radial region of the wave functions, respectively. Therefore, we defined a radial dependent factor $C_{if}(r)$ by
\begin{eqnarray}
\label{eqn}
{M_{if}} = \int_0^\infty  {{C_{if}(r)}dr},
\end{eqnarray}
where the $M_{if}$ is the radiative transition matrix element. In Fig.~\ref{gauge}, we illustrated the radial dependence of $C_{if}(r)$ for the transition matrix element $\langle^7F_0|O^{(1)}|^9G^{o}_1\rangle$ in Babushkin and Coulomb gauges.
It was found that only the wave function in the larger $r$ region contribute significantly to the $E1$ transition amplitude in Babushkin gauge, while the transition matrix element in Coulomb gauge are very sensitive to the whole region of wave functions. Meanwhile, there are large cancellations in the integral of transition matrix element in coulomb gauges, which lead to the requirement of higher quality wave functions. As a result, the oscillator strengths in the Babushkin gauge is more reliable than ones in the Coulomb gauge.

In this work, only the selected first-order electron correlations were taken into account in the calculations. The uncertainties of the transition energies and oscillator strengths are mainly attributed to the higher-order and the residual first-order electron correlations. 
The comparison with the experimental data presented in Table~\ref{tab:table9} can give us a rough estimate of the uncertainties of our results. Approximately, the errors in present calculations of the transition energies are 2\%-3\%. For the oscillator strengths, the differences between the results in Babushkin gauge and experimental values are about 30\% for the transition $^9G^{o}_1$~-~$^7F_{2}$, and 10\% for the stronger lines. Next, we examined the influence of different correlation models on the HFS constants.



\subsection{ \label{sec:HFS} The hyperfine constants }

\begin{figure*}
\subfigure [from different one-body electron correlations]{
\includegraphics[width=0.5\textwidth]{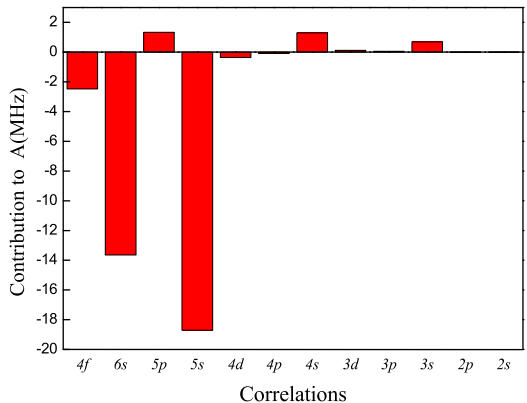}}\subfigure [from different two-body electron correlations]{
\includegraphics[width=0.5\textwidth]{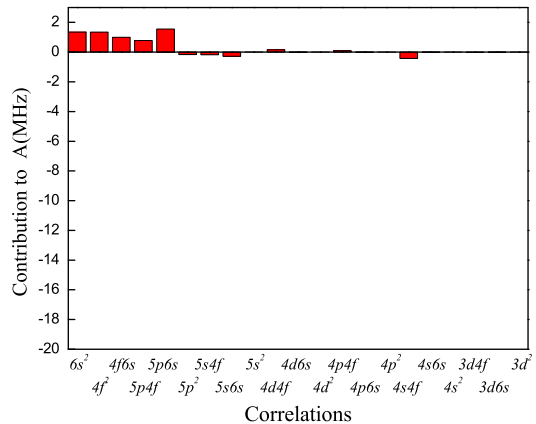}}

\subfigure [from different one-body electron correlations]{
\includegraphics[width=0.5\textwidth]{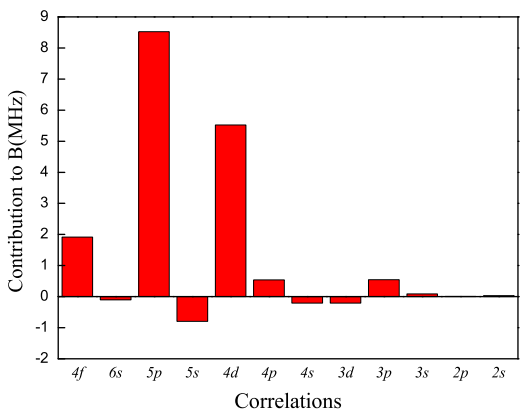}}\subfigure [from different two-body electron correlations]{
\includegraphics[width=0.5\textwidth]{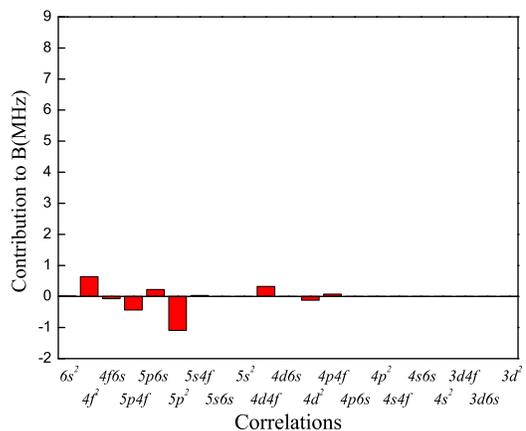}}
\caption{\label{fig:6}(Color online) Contributions from different one- and two-body electron correlation effects to HFS constants $A$ and $B$ of the $4f^{6}6s^{2}$ $^{7}F_{1}$ state. The contributions were evaluated by the difference between the multi- and single-configuration values.}
\end{figure*}

Atomic HFS provides important test for \emph{ab initio} atomic-structure calculation, since hyperfine interactions are sensitive to electron correlations. In 1985, Cheng and Childs calculated the ground state multiplet HFS constants of Sm atom within the single-configuration approximation~\cite{Cheng1985}.
The results are in quite good agreement with experiment~\cite{Childs2011}. Latter, a multi-configuration calculation was reported in 2001 by Dilip \emph{et al.}~\cite{Dilip2001} with a worse results compared with experimental values. In view of this, we investigated different electron correlation effects on HFS constants, and then carried out new multi-configuration calculations for $4f^{6}6s^{2}$ $^{7}F_{1}$ state.

\begin{table}
\caption{\label{tab:table10}%
The hyperfine constants $A$, $B$ and $B/Q$ for $^{7}F_{1}$ state of $^{147}$Sm. `VV' = multi-configurations calculation with valence-valence correlations; ``n$spdfg$'' = virtual orbital set of the MCDHF calculation with the correlation model described in Sec.~\ref{sec:HFS}; $\mu = -0.812~\mu _N$ and $Q = -0.261~b$ were taken from the Ref~\cite{Stone2005}.
}
\begin{ruledtabular}
\begin{tabular}{ccccc}
Model                         & $A$(MHz) & $B$(MHz) & $B/Q$(MHz/b)\\
\hline
single-configuration          & -33.73 & -75.89 & 290.77 \\
 ``$1spdfg$''                 & -99.45 & -26.22 & 100.46 \\
 ``$2spdfg$''                 & -30.77 & -60.19 & 230.61 \\
 ``$3spdfg$''                 & -31.89 & -60.88 & 233.26 \\
A. Dilip\cite{Dilip2001}      & -23.12 & -35.70 &        \\
K. T. Cheng\cite{Cheng1985}   & -33.77 & -58.88\footnotemark[1] & 290.05 \\
Experiment\cite{Childs2011}   & -33.493& -58.688& 224.86 \\
\end{tabular}
\end{ruledtabular}
\footnotetext[1]{calculated with $Q = -0.203b$}
\end{table}

\begin{table}
\caption{\label{tab:table11}%
The hyperfine constants $A$ and $B$ for $4f^66s6p$ $^{9}F^{o}_{1}$ and $^{7}G^{o}_{1}$ state of $^{147}$Sm in units of MHz; ``n$spdf$'' = virtual orbital set.
}
\begin{ruledtabular}
\begin{tabular}{crrrr}
                           & \multicolumn{2}{c}{$^{9}F^{o}_{1}$} & \multicolumn{2}{c}{$^{7}G^{o}_{1}$}\\
model                      & A(MHz)  & B(MHz)                    & A(MHz)  & B(MHz) \\
\hline
single-configuration       & -241.05 & 19.24                     & -171.83 & 10.79 \\
``$1spdf$''                    & -539.05 & 17.27                     & -89.22  & -14.11 \\
``$2spdf$''                    & -326.12 & 15.86                     & -169.75 & -12.49   \\
Experiment~\cite{Park2003} & -423.34 & 13.21                     & -212.62 & -9.63     \\
\end{tabular}
\end{ruledtabular}
\end{table}

In Fig.~\ref{fig:6}, we presented the contributions from the different correlations to the magnetic dipole HFS constants $A$ and electric quadrupole constant $B$ of the $4f^{6}6s^{2}$ $^{7}F_{1}$ state for $^{147}$Sm isotope with nuclear spin $I = 7/2$. Using the experimental nuclear parameters taken from the Ref~\cite{Stone2005}, the contributions were evaluated by the difference between the multi- and single-configuration values. It was found that the one-body electron correlations are very important to the constants, since the major corrections for hyperfine interaction are the spin and orbital polarizations~\cite{Fischer1997}. The two-body electron correlations have a relatively smaller effect on the HFS constants, although essentially important for the total energy.
Therefore, in our MCDHF calculations the one-body electron correlations from $3spd4spdf5sp6s$ shells were chosen to form atomic state wave functions.

Using the computational model described above, the HFS constants $A$, $B$ and $B/Q$ of the $4f^{6}6s^{2}$ $^{7}F_{1}$ state for $^{147}$Sm isotope were presented in Table~\ref{tab:table10}.  In the single-configuration approximation, our result of the constant $A$ agrees with the similar work of Cheng and Childs~\cite{Cheng1985} and experimental measurement~\cite{Childs2011} quite well. Ref~\cite{Cheng1985} provide a better result of constant $B$, but they used a fitting electric quadrupole moment $Q$. The similar $B/Q$ results from Ref~\cite{Cheng1985} and ours were given in this table.
In our multi-configuration calculations, the HFS constants $A$ and $B$ were significantly changed by the considered electron correlations, and then were converged with the expansion of the configuration space. The results of both constants A and B have a good agreement with experimental data. Moreover, the constant $B$ was improved from the single-configuration calculation, only about 2 MHz lower than the experimental value. The large discrepancies between Dilip's multi-configuration calculation~\cite{Dilip2001} and experimental data could be due to that they only partly considered VV correlations, which couldn't improve the results when the electron correlation effects balance out.

For exited state, there are no theoretical predictions of HFS according to the best of our knowledge. Using the same computational model, the results of the HFS constants for $4f^66s6p$ $^{9}F^{o}_{1}$ and $^{7}G^{o}_{1}$ states were shown in table~\ref{tab:table11}.  
In this case, the single-configuration calculations couldn't provide reasonable results. For example, the calculated HFS constant $B$ of $^{7}G^o_{1}$ state is 10.79 MHz while the experimental value is -9.63~\cite{Park2003}. Although our MCDHF calculations of HFS constant $A$ were not fully converged, the values of constant $B$ were largely improved by the captured one-body correlation effects. The constant $B$ of $^{7}G^o_{1}$ state becomes -12.49, much closer to the experimental value.

In our calculations of hyperfine constants, only the one-body electron correlations were considered. All the two two-body correlations and the higher-order corrections contribute to the uncertainties. However, the errors of both the constants $A$ and $B$ of the $4f^{6}6s^{2}$ $^{7}F_{1}$ state were found to be less than 5\%. For the $4f^66s6p$ $^{9}F^{o}_{1}$ and $^{7}G^{o}_{1}$ states, the accuracy of the results mainly depended on the convergence of the calculations. Comparing with experimental values, the deviation of the hyperfine constants of $^{9}F^{o}_{1}$ and $^{7}G^{o}_{1}$ states are about 20\%-30\%.

\section{CONCLUSION}

Recently, partitioned correlation function interaction (PCFI) approach was developed for complicated atoms \cite{FroeseFischer2013,Verdebout2013}, which relax the orthonormality restriction on the orbital basis and break down the originally very large calculations into a series of smaller calculations. The subspace of CFSs makes it easier to capture the effects weakly connected to total energy, which could be significant for some atomic properties. For the calculations of complicated atom like Sm, with complicated and strong electron correlation effects, it would be very useful to divide the electron correlations. In this work, we also divided the first-order electron correlations into several subsets, but used the multi-configuration Dirac-Hartree-Fock (MCDHF) method to investigated these correlation effects on total energies, excitation energies and HFS constants. It was found that the core correlations are of importance for the total energies. However, only the correlations involving $6s$ and $6p$ valence orbitals significantly influence the excitation energies, although they make relatively small contributions to total energies. For HFS constants, the major corrections are from the one-body electron correlation effects.

Based on the analysis of electron correlation effects, the important configuration state wave functions were selected to calculate the different atomic properties using the MCDHF approach. The results of transition energies and oscillator strengths from $4f^{6}6s6p$ $^{9}G_{1}^{o}$ and $^{9}F_{1}^{o}$ to $4f^{6}6s^{2}$ $^{7}F_{0,1,2}$ states have a much better agreement with experiment, compared with previous calculations without core correlations. 
Furthermore, the HFS constants were also calculated for examining correlation models. 
It was found that the validity of single configuration approximation is restricted on ground state multiplet and more complicated electron correlations are requited for treating the HFS for excited states.
By including the important correlation effects, the reasonable results were obtained for $4f^66s6p$ $^{9}F^{o}_{1}$, $^{7}G^{o}_{1}$ and $4f^66s^2$ $^7F_1$ states.
%


\nocite{*}

\acknowledgments{We would like to thank Prof. Per J\"{o}nsson for his helpful discussions. This work was supported by the NSAF (Grants No. U1330117), National Natural Science Foundation of China (Grant No. 11404025) and China Postdoctoral Science Foundation (Grant No.2014M560061). }
\bibliography{samarium}

\begin{thebibliography}{31}%
\makeatletter
\providecommand \@ifxundefined [1]{%
 \@ifx{#1\undefined}
}%
\providecommand \@ifnum [1]{%
 \ifnum #1\expandafter \@firstoftwo
 \else \expandafter \@secondoftwo
 \fi
}%
\providecommand \@ifx [1]{%
 \ifx #1\expandafter \@firstoftwo
 \else \expandafter \@secondoftwo
 \fi
}%
\providecommand \natexlab [1]{#1}%
\providecommand \enquote  [1]{``#1''}%
\providecommand \bibnamefont  [1]{#1}%
\providecommand \bibfnamefont [1]{#1}%
\providecommand \citenamefont [1]{#1}%
\providecommand \href@noop [0]{\@secondoftwo}%
\providecommand \href [0]{\begingroup \@sanitize@url \@href}%
\providecommand \@href[1]{\@@startlink{#1}\@@href}%
\providecommand \@@href[1]{\endgroup#1\@@endlink}%
\providecommand \@sanitize@url [0]{\catcode `\\12\catcode `\$12\catcode
  `\&12\catcode `\#12\catcode `\^12\catcode `\_12\catcode `\%12\relax}%
\providecommand \@@startlink[1]{}%
\providecommand \@@endlink[0]{}%
\providecommand \url  [0]{\begingroup\@sanitize@url \@url }%
\providecommand \@url [1]{\endgroup\@href {#1}{\urlprefix }}%
\providecommand \urlprefix  [0]{URL }%
\providecommand \Eprint [0]{\href }%
\providecommand \doibase [0]{http://dx.doi.org/}%
\providecommand \selectlanguage [0]{\@gobble}%
\providecommand \bibinfo  [0]{\@secondoftwo}%
\providecommand \bibfield  [0]{\@secondoftwo}%
\providecommand \translation [1]{[#1]}%
\providecommand \BibitemOpen [0]{}%
\providecommand \bibitemStop [0]{}%
\providecommand \bibitemNoStop [0]{.\EOS\space}%
\providecommand \EOS [0]{\spacefactor3000\relax}%
\providecommand \BibitemShut  [1]{\csname bibitem#1\endcsname}%
\let\auto@bib@innerbib\@empty
\bibitem [{\citenamefont {B\"{u}nzli}(2010)}]{Bunzli2010}%
  \BibitemOpen
  \bibfield  {author} {\bibinfo {author} {\bibfnamefont {J.-C.~G.}\
  \bibnamefont {B\"{u}nzli}},\ }\href {\doibase 10.1021/cr900362e} {\bibfield
  {journal} {\bibinfo  {journal} {Chemical Reviews}\ }\textbf {\bibinfo
  {volume} {110}},\ \bibinfo {pages} {2729} (\bibinfo {year}
  {2010})}\BibitemShut {NoStop}%
\bibitem [{\citenamefont {Lawler}\ \emph {et~al.}(2009)\citenamefont {Lawler},
  \citenamefont {Sneden}, \citenamefont {Cowan}, \citenamefont {Ivans},\ and\
  \citenamefont {{Den Hartog}}}]{Lawler2009}%
  \BibitemOpen
  \bibfield  {author} {\bibinfo {author} {\bibfnamefont {J.~E.}\ \bibnamefont
  {Lawler}}, \bibinfo {author} {\bibfnamefont {C.}~\bibnamefont {Sneden}},
  \bibinfo {author} {\bibfnamefont {J.~J.}\ \bibnamefont {Cowan}}, \bibinfo
  {author} {\bibfnamefont {I.~I.}\ \bibnamefont {Ivans}}, \ and\ \bibinfo
  {author} {\bibfnamefont {E.~a.}\ \bibnamefont {{Den Hartog}}},\ }\href
  {\doibase 10.1088/0067-0049/182/1/51} {\bibfield  {journal} {\bibinfo
  {journal} {The Astrophysical Journal Supplement Series}\ }\textbf {\bibinfo
  {volume} {182}},\ \bibinfo {pages} {51} (\bibinfo {year} {2009})}\BibitemShut
  {NoStop}%
\bibitem [{\citenamefont {Sneden}\ \emph {et~al.}(2009)\citenamefont {Sneden},
  \citenamefont {Lawler}, \citenamefont {Cowan}, \citenamefont {Ivans},\ and\
  \citenamefont {{Den Hartog}}}]{Sneden2009}%
  \BibitemOpen
  \bibfield  {author} {\bibinfo {author} {\bibfnamefont {C.}~\bibnamefont
  {Sneden}}, \bibinfo {author} {\bibfnamefont {J.~E.}\ \bibnamefont {Lawler}},
  \bibinfo {author} {\bibfnamefont {J.~J.}\ \bibnamefont {Cowan}}, \bibinfo
  {author} {\bibfnamefont {I.~I.}\ \bibnamefont {Ivans}}, \ and\ \bibinfo
  {author} {\bibfnamefont {E.~a.}\ \bibnamefont {{Den Hartog}}},\ }\href
  {\doibase 10.1088/0067-0049/182/1/80} {\bibfield  {journal} {\bibinfo
  {journal} {The Astrophysical Journal Supplement Series}\ }\textbf {\bibinfo
  {volume} {182}},\ \bibinfo {pages} {80} (\bibinfo {year} {2009})}\BibitemShut
  {NoStop}%
\bibitem [{\citenamefont {B\"{u}nzli}\ and\ \citenamefont
  {Piguet}(2005)}]{B406082M}%
  \BibitemOpen
  \bibfield  {author} {\bibinfo {author} {\bibfnamefont {J.-C.~G.}\
  \bibnamefont {B\"{u}nzli}}\ and\ \bibinfo {author} {\bibfnamefont
  {C.}~\bibnamefont {Piguet}},\ }\href {\doibase 10.1039/B406082M} {\bibfield
  {journal} {\bibinfo  {journal} {Chem. Soc. Rev.}\ }\textbf {\bibinfo {volume}
  {34}},\ \bibinfo {pages} {1048} (\bibinfo {year} {2005})}\BibitemShut
  {NoStop}%
\bibitem [{\citenamefont {Martin}\ \emph {et~al.}(1978)\citenamefont {Martin},
  \citenamefont {Zalubas},\ and\ \citenamefont {Hagan}}]{Martin1978}%
  \BibitemOpen
  \bibfield  {author} {\bibinfo {author} {\bibfnamefont {W.~C.}\ \bibnamefont
  {Martin}}, \bibinfo {author} {\bibfnamefont {R.}~\bibnamefont {Zalubas}}, \
  and\ \bibinfo {author} {\bibfnamefont {L.}~\bibnamefont {Hagan}},\
  }\href@noop {} {\emph {\bibinfo {title} {{Atomic Energy Levels—The
  Rare-Earth Elements}}}}\ (\bibinfo  {publisher} {Natl. Bur. Stand. Ref. Data
  Se., Natl. Bur. Stand. (U.S.), NBS-60,422},\ \bibinfo {year}
  {1978})\BibitemShut {NoStop}%
\bibitem [{\citenamefont {Beck}\ and\ \citenamefont
  {Domeier}(2009)}]{Beck2009}%
  \BibitemOpen
  \bibfield  {author} {\bibinfo {author} {\bibfnamefont {D.~R.}\ \bibnamefont
  {Beck}}\ and\ \bibinfo {author} {\bibfnamefont {E.}~\bibnamefont {Domeier}},\
  }\href {\doibase 10.1139/P08-051} {\bibfield  {journal} {\bibinfo  {journal}
  {Can. J. Phys.}\ }\textbf {\bibinfo {volume} {81}},\ \bibinfo {pages} {75}
  (\bibinfo {year} {2009})}\BibitemShut {NoStop}%
\bibitem [{\citenamefont {Beck}\ and\ \citenamefont
  {O'Malley}(2010)}]{Beck2010}%
  \BibitemOpen
  \bibfield  {author} {\bibinfo {author} {\bibfnamefont {D.~R.}\ \bibnamefont
  {Beck}}\ and\ \bibinfo {author} {\bibfnamefont {S.~M.}\ \bibnamefont
  {O'Malley}},\ }\href {\doibase 10.1088/0953-4075/43/21/215003} {\bibfield
  {journal} {\bibinfo  {journal} {Journal of Physics B}\ }\textbf {\bibinfo
  {volume} {43}},\ \bibinfo {pages} {215003} (\bibinfo {year}
  {2010})}\BibitemShut {NoStop}%
\bibitem [{\citenamefont {Kramida}\ \emph {et~al.}(2014)\citenamefont
  {Kramida}, \citenamefont {{Yu.~Ralchenko}}, \citenamefont {Reader},\ and\
  \citenamefont {{NIST ASD Team}}}]{NIST_ASD}%
  \BibitemOpen
  \bibfield  {author} {\bibinfo {author} {\bibfnamefont {A.}~\bibnamefont
  {Kramida}}, \bibinfo {author} {\bibnamefont {{Yu.~Ralchenko}}}, \bibinfo
  {author} {\bibfnamefont {J.}~\bibnamefont {Reader}}, \ and\ \bibinfo {author}
  {\bibnamefont {{NIST ASD Team}}},\ }\href@noop {} {}\bibinfo {howpublished}
  {{NIST Atomic Spectra Database (ver. 5.2), [Online]. Available:
  {\tt{http://physics.nist.gov/asd}} [2014, September 22]. National Institute
  of Standards and Technology, Gaithersburg, MD.}} (\bibinfo {year}
  {2014})\BibitemShut {NoStop}%
\bibitem [{\citenamefont {Porsev}(1997)}]{Porsev1997}%
  \BibitemOpen
  \bibfield  {author} {\bibinfo {author} {\bibfnamefont {S.}~\bibnamefont
  {Porsev}},\ }\href {\doibase 10.1103/PhysRevA.56.3535} {\bibfield  {journal}
  {\bibinfo  {journal} {Physical Review A}\ }\textbf {\bibinfo {volume} {56}},\
  \bibinfo {pages} {3535} (\bibinfo {year} {1997})}\BibitemShut {NoStop}%
\bibitem [{\citenamefont {Blagoev}\ \emph {et~al.}(1977)\citenamefont
  {Blagoev}, \citenamefont {Komarovskii},\ and\ \citenamefont
  {Penkin}}]{Blagoev1977}%
  \BibitemOpen
  \bibfield  {author} {\bibinfo {author} {\bibfnamefont {K.~B.}\ \bibnamefont
  {Blagoev}}, \bibinfo {author} {\bibfnamefont {V.~A.}\ \bibnamefont
  {Komarovskii}}, \ and\ \bibinfo {author} {\bibfnamefont {N.~P.}\ \bibnamefont
  {Penkin}},\ }\href@noop {} {\bibfield  {journal} {\bibinfo  {journal} {Opt.
  Spektrosk}\ }\textbf {\bibinfo {volume} {42}},\ \bibinfo {pages} {424}
  (\bibinfo {year} {1977})}\BibitemShut {NoStop}%
\bibitem [{\citenamefont {Lawler}\ \emph {et~al.}(2013)\citenamefont {Lawler},
  \citenamefont {Fittante},\ and\ \citenamefont {{Den Hartog}}}]{Lawler2013}%
  \BibitemOpen
  \bibfield  {author} {\bibinfo {author} {\bibfnamefont {J.~E.}\ \bibnamefont
  {Lawler}}, \bibinfo {author} {\bibfnamefont {a.~J.}\ \bibnamefont
  {Fittante}}, \ and\ \bibinfo {author} {\bibfnamefont {E.~a.}\ \bibnamefont
  {{Den Hartog}}},\ }\href {\doibase 10.1088/0953-4075/46/21/215004} {\bibfield
   {journal} {\bibinfo  {journal} {Journal of Physics B}\ }\textbf {\bibinfo
  {volume} {46}},\ \bibinfo {pages} {215004} (\bibinfo {year}
  {2013})}\BibitemShut {NoStop}%
\bibitem [{\citenamefont {Komarovskii}\ and\ \citenamefont
  {Smirnov}(1996)}]{Komarovskii1996}%
  \BibitemOpen
  \bibfield  {author} {\bibinfo {author} {\bibfnamefont {V.~A.}\ \bibnamefont
  {Komarovskii}}\ and\ \bibinfo {author} {\bibfnamefont {Y.~M.}\ \bibnamefont
  {Smirnov}},\ }\href@noop {} {\bibfield  {journal} {\bibinfo  {journal} {Opt.
  Spektrosk.}\ }\textbf {\bibinfo {volume} {80}},\ \bibinfo {pages} {357}
  (\bibinfo {year} {1996})}\BibitemShut {NoStop}%
\bibitem [{\citenamefont {Komarovskii}\ \emph {et~al.}(1970)\citenamefont
  {Komarovskii}, \citenamefont {Penkin},\ and\ \citenamefont
  {Nikiforova}}]{Komarovskii1970}%
  \BibitemOpen
  \bibfield  {author} {\bibinfo {author} {\bibfnamefont {V.~A.}\ \bibnamefont
  {Komarovskii}}, \bibinfo {author} {\bibfnamefont {N.~P.}\ \bibnamefont
  {Penkin}}, \ and\ \bibinfo {author} {\bibfnamefont {G.~P.}\ \bibnamefont
  {Nikiforova}},\ }\href@noop {} {\bibfield  {journal} {\bibinfo  {journal}
  {Opt. Spektrosk}\ }\textbf {\bibinfo {volume} {29}},\ \bibinfo {pages} {220}
  (\bibinfo {year} {1970})}\BibitemShut {NoStop}%
\bibitem [{\citenamefont {Dilip}\ \emph {et~al.}(2001)\citenamefont {Dilip},
  \citenamefont {Endo}, \citenamefont {Fukumi}, \citenamefont {Iinuma},
  \citenamefont {Kondo},\ and\ \citenamefont {Takahashi}}]{Dilip2001}%
  \BibitemOpen
  \bibfield  {author} {\bibinfo {author} {\bibfnamefont {A.}~\bibnamefont
  {Dilip}}, \bibinfo {author} {\bibfnamefont {I.}~\bibnamefont {Endo}},
  \bibinfo {author} {\bibfnamefont {A.}~\bibnamefont {Fukumi}}, \bibinfo
  {author} {\bibfnamefont {M.}~\bibnamefont {Iinuma}}, \bibinfo {author}
  {\bibfnamefont {T.}~\bibnamefont {Kondo}}, \ and\ \bibinfo {author}
  {\bibfnamefont {T.}~\bibnamefont {Takahashi}},\ }\href@noop {} {\bibfield
  {journal} {\bibinfo  {journal} {Eur. Phys. J. D}\ }\textbf {\bibinfo {volume}
  {277}},\ \bibinfo {pages} {271} (\bibinfo {year} {2001})}\BibitemShut
  {NoStop}%
\bibitem [{\citenamefont {Childs}\ and\ \citenamefont
  {Goodman}(1972)}]{Childs2011}%
  \BibitemOpen
  \bibfield  {author} {\bibinfo {author} {\bibfnamefont {W.}~\bibnamefont
  {Childs}}\ and\ \bibinfo {author} {\bibfnamefont {L.}~\bibnamefont
  {Goodman}},\ }\href {\doibase 10.1103/PhysRevA.6.2011} {\bibfield  {journal}
  {\bibinfo  {journal} {Physical Review A}\ }\textbf {\bibinfo {volume} {6}},\
  \bibinfo {pages} {2011} (\bibinfo {year} {1972})}\BibitemShut {NoStop}%
\bibitem [{\citenamefont {Cheng}\ and\ \citenamefont
  {Childs}(1985)}]{Cheng1985}%
  \BibitemOpen
  \bibfield  {author} {\bibinfo {author} {\bibfnamefont {K.~T.}\ \bibnamefont
  {Cheng}}\ and\ \bibinfo {author} {\bibfnamefont {W.~J.}\ \bibnamefont
  {Childs}},\ }\href@noop {} {\bibfield  {journal} {\bibinfo  {journal}
  {Physical Review A}\ }\textbf {\bibinfo {volume} {31}},\ \bibinfo {pages}
  {2775} (\bibinfo {year} {1985})}\BibitemShut {NoStop}%
\bibitem [{\citenamefont {Zou}\ and\ \citenamefont {{Froese
  Fischer}}(2002)}]{zou2002}%
  \BibitemOpen
  \bibfield  {author} {\bibinfo {author} {\bibfnamefont {Y.}~\bibnamefont
  {Zou}}\ and\ \bibinfo {author} {\bibfnamefont {C.}~\bibnamefont {{Froese
  Fischer}}},\ }\href {\doibase 10.1103/PhysRevLett.88.183001} {\bibfield
  {journal} {\bibinfo  {journal} {Physical Review Letters}\ }\textbf {\bibinfo
  {volume} {88}},\ \bibinfo {pages} {183001} (\bibinfo {year}
  {2002})}\BibitemShut {NoStop}%
\bibitem [{\citenamefont {{I. P. Grant}}(2007)}]{I.P.Grant2007}%
  \BibitemOpen
  \bibfield  {author} {\bibinfo {author} {\bibnamefont {{I. P. Grant}}},\
  }\href@noop {} {\emph {\bibinfo {title} {{Relativistic Quantum Theory of
  Atoms and Molecules}}}}\ (\bibinfo  {publisher} {Springer},\ \bibinfo
  {address} {New York},\ \bibinfo {year} {2007})\BibitemShut {NoStop}%
\bibitem [{\citenamefont {Fischer}\ \emph {et~al.}(1997)\citenamefont
  {Fischer}, \citenamefont {Brage},\ and\ \citenamefont
  {J\"{o}nsson}}]{Fischer1997}%
  \BibitemOpen
  \bibfield  {author} {\bibinfo {author} {\bibfnamefont {C.~F.}\ \bibnamefont
  {Fischer}}, \bibinfo {author} {\bibfnamefont {T.}~\bibnamefont {Brage}}, \
  and\ \bibinfo {author} {\bibfnamefont {P.}~\bibnamefont {J\"{o}nsson}},\
  }\href@noop {} {\emph {\bibinfo {title} {{Computational Atomic Structure An
  MCHF Approach}}}}\ (\bibinfo  {publisher} {Institute of Physics Publishing},\
  \bibinfo {address} {London},\ \bibinfo {year} {1997})\BibitemShut {NoStop}%
\bibitem [{\citenamefont {J\"{o}nsson}\ \emph {et~al.}(2013)\citenamefont
  {J\"{o}nsson}, \citenamefont {Gaigalas}, \citenamefont {Biero\'{n}},
  \citenamefont {{Froese Fischer}},\ and\ \citenamefont {Grant}}]{Jonsson2013}%
  \BibitemOpen
  \bibfield  {author} {\bibinfo {author} {\bibfnamefont {P.}~\bibnamefont
  {J\"{o}nsson}}, \bibinfo {author} {\bibfnamefont {G.}~\bibnamefont
  {Gaigalas}}, \bibinfo {author} {\bibfnamefont {J.}~\bibnamefont
  {Biero\'{n}}}, \bibinfo {author} {\bibfnamefont {C.}~\bibnamefont {{Froese
  Fischer}}}, \ and\ \bibinfo {author} {\bibfnamefont {I.}~\bibnamefont
  {Grant}},\ }\href {\doibase 10.1016/j.cpc.2013.02.016} {\bibfield  {journal}
  {\bibinfo  {journal} {Computer Physics Communications}\ }\textbf {\bibinfo
  {volume} {184}},\ \bibinfo {pages} {2197} (\bibinfo {year}
  {2013})}\BibitemShut {NoStop}%
\bibitem [{\citenamefont {J\"{o}nsson}\ \emph {et~al.}(2007)\citenamefont
  {J\"{o}nsson}, \citenamefont {He}, \citenamefont {{Froese Fischer}},\ and\
  \citenamefont {Grant}}]{Jonsson2007}%
  \BibitemOpen
  \bibfield  {author} {\bibinfo {author} {\bibfnamefont {P.}~\bibnamefont
  {J\"{o}nsson}}, \bibinfo {author} {\bibfnamefont {X.}~\bibnamefont {He}},
  \bibinfo {author} {\bibfnamefont {C.}~\bibnamefont {{Froese Fischer}}}, \
  and\ \bibinfo {author} {\bibfnamefont {I.}~\bibnamefont {Grant}},\ }\href
  {\doibase 10.1016/j.cpc.2007.06.002} {\bibfield  {journal} {\bibinfo
  {journal} {Computer Physics Communications}\ }\textbf {\bibinfo {volume}
  {177}},\ \bibinfo {pages} {597} (\bibinfo {year} {2007})}\BibitemShut
  {NoStop}%
\bibitem [{\citenamefont {Parpia}\ \emph {et~al.}(1996)\citenamefont {Parpia},
  \citenamefont {{Froese Fischer}},\ and\ \citenamefont {Grant}}]{Parpia1996}%
  \BibitemOpen
  \bibfield  {author} {\bibinfo {author} {\bibfnamefont {F.}~\bibnamefont
  {Parpia}}, \bibinfo {author} {\bibfnamefont {C.}~\bibnamefont {{Froese
  Fischer}}}, \ and\ \bibinfo {author} {\bibfnamefont {I.}~\bibnamefont
  {Grant}},\ }\href {\doibase 10.1016/0010-4655(95)00136-0} {\bibfield
  {journal} {\bibinfo  {journal} {Computer Physics Communications}\ }\textbf
  {\bibinfo {volume} {94}},\ \bibinfo {pages} {249} (\bibinfo {year}
  {1996})}\BibitemShut {NoStop}%
\bibitem [{\citenamefont {Fano}(1965)}]{Fano1965}%
  \BibitemOpen
  \bibfield  {author} {\bibinfo {author} {\bibfnamefont {U.}~\bibnamefont
  {Fano}},\ }\href@noop {} {\bibfield  {journal} {\bibinfo  {journal} {Physical
  Review A}\ }\textbf {\bibinfo {volume} {140}},\ \bibinfo {pages} {67}
  (\bibinfo {year} {1965})}\BibitemShut {NoStop}%
\bibitem [{\citenamefont {Olsen}\ \emph {et~al.}(1995)\citenamefont {Olsen},
  \citenamefont {Godefroid}, \citenamefont {J\"{o}nsson}, \citenamefont
  {Malmqvist},\ and\ \citenamefont {Froese}}]{Olsen1995}%
  \BibitemOpen
  \bibfield  {author} {\bibinfo {author} {\bibfnamefont {J.}~\bibnamefont
  {Olsen}}, \bibinfo {author} {\bibfnamefont {M.~R.}\ \bibnamefont
  {Godefroid}}, \bibinfo {author} {\bibfnamefont {P.}~\bibnamefont
  {J\"{o}nsson}}, \bibinfo {author} {\bibfnamefont {P.~A.}\ \bibnamefont
  {Malmqvist}}, \ and\ \bibinfo {author} {\bibfnamefont {F.~C.}\ \bibnamefont
  {Froese}},\ }\href@noop {} {\bibfield  {journal} {\bibinfo  {journal}
  {Physical Review E}\ }\textbf {\bibinfo {volume} {52}},\ \bibinfo {pages}
  {4499} (\bibinfo {year} {1995})}\BibitemShut {NoStop}%
\bibitem [{\citenamefont {J\"{o}nsson}\ \emph {et~al.}(1996)\citenamefont
  {J\"{o}nsson}, \citenamefont {Parpia},\ and\ \citenamefont {{Froese
  Fischer}}}]{Jonsson1996}%
  \BibitemOpen
  \bibfield  {author} {\bibinfo {author} {\bibfnamefont {P.}~\bibnamefont
  {J\"{o}nsson}}, \bibinfo {author} {\bibfnamefont {F.}~\bibnamefont {Parpia}},
  \ and\ \bibinfo {author} {\bibfnamefont {C.}~\bibnamefont {{Froese
  Fischer}}},\ }\href@noop {} {\bibfield  {journal} {\bibinfo  {journal}
  {Computer Physics Communications}\ }\textbf {\bibinfo {volume} {96}},\
  \bibinfo {pages} {301} (\bibinfo {year} {1996})}\BibitemShut {NoStop}%
\bibitem [{\citenamefont {Li}\ \emph {et~al.}(2012)\citenamefont {Li},
  \citenamefont {J\"{o}nsson}, \citenamefont {Godefroid}, \citenamefont
  {Dong},\ and\ \citenamefont {Gaigalas}}]{Li2012}%
  \BibitemOpen
  \bibfield  {author} {\bibinfo {author} {\bibfnamefont {J.~G.}\ \bibnamefont
  {Li}}, \bibinfo {author} {\bibfnamefont {P.}~\bibnamefont {J\"{o}nsson}},
  \bibinfo {author} {\bibfnamefont {M.}~\bibnamefont {Godefroid}}, \bibinfo
  {author} {\bibfnamefont {C.~Z.}\ \bibnamefont {Dong}}, \ and\ \bibinfo
  {author} {\bibfnamefont {G.}~\bibnamefont {Gaigalas}},\ }\href {\doibase
  10.1103/PhysRevA.86.052523} {\bibfield  {journal} {\bibinfo  {journal}
  {Physical Review A}\ }\textbf {\bibinfo {volume} {86}},\ \bibinfo {pages}
  {052523} (\bibinfo {year} {2012})}\BibitemShut {NoStop}%
\bibitem [{\citenamefont {Biero\'{n}}\ \emph {et~al.}(2009)\citenamefont
  {Biero\'{n}}, \citenamefont {{Froese Fischer}}, \citenamefont {Indelicato},
  \citenamefont {J\"{o}nsson},\ and\ \citenamefont {Pyykk\"{o}}}]{Bieron2009}%
  \BibitemOpen
  \bibfield  {author} {\bibinfo {author} {\bibfnamefont {J.}~\bibnamefont
  {Biero\'{n}}}, \bibinfo {author} {\bibfnamefont {C.}~\bibnamefont {{Froese
  Fischer}}}, \bibinfo {author} {\bibfnamefont {P.}~\bibnamefont {Indelicato}},
  \bibinfo {author} {\bibfnamefont {P.}~\bibnamefont {J\"{o}nsson}}, \ and\
  \bibinfo {author} {\bibfnamefont {P.}~\bibnamefont {Pyykk\"{o}}},\ }\href
  {\doibase 10.1103/PhysRevA.79.052502} {\bibfield  {journal} {\bibinfo
  {journal} {Physical Review A}\ }\textbf {\bibinfo {volume} {79}},\ \bibinfo
  {pages} {052502} (\bibinfo {year} {2009})}\BibitemShut {NoStop}%
\bibitem [{\citenamefont {Stone}(2005)}]{Stone2005}%
  \BibitemOpen
  \bibfield  {author} {\bibinfo {author} {\bibfnamefont {N.}~\bibnamefont
  {Stone}},\ }\href {\doibase 10.1016/j.adt.2005.04.001} {\bibfield  {journal}
  {\bibinfo  {journal} {Atomic Data and Nuclear Data Tables}\ }\textbf
  {\bibinfo {volume} {90}},\ \bibinfo {pages} {75} (\bibinfo {year}
  {2005})}\BibitemShut {NoStop}%
\bibitem [{\citenamefont {Park}\ \emph {et~al.}(2003)\citenamefont {Park},
  \citenamefont {Lee},\ and\ \citenamefont {Rhee}}]{Park2003}%
  \BibitemOpen
  \bibfield  {author} {\bibinfo {author} {\bibfnamefont {H.}~\bibnamefont
  {Park}}, \bibinfo {author} {\bibfnamefont {M.}~\bibnamefont {Lee}}, \ and\
  \bibinfo {author} {\bibfnamefont {Y.}~\bibnamefont {Rhee}},\ }\href@noop {}
  {\bibfield  {journal} {\bibinfo  {journal} {Journal of the Korean Physical
  Society}\ }\textbf {\bibinfo {volume} {43}},\ \bibinfo {pages} {336}
  (\bibinfo {year} {2003})}\BibitemShut {NoStop}%
\bibitem [{\citenamefont {{Froese Fischer}}\ \emph {et~al.}(2013)\citenamefont
  {{Froese Fischer}}, \citenamefont {Verdebout}, \citenamefont {Godefroid},
  \citenamefont {Rynkun}, \citenamefont {J\"{o}nsson},\ and\ \citenamefont
  {Gaigalas}}]{FroeseFischer2013}%
  \BibitemOpen
  \bibfield  {author} {\bibinfo {author} {\bibfnamefont {C.}~\bibnamefont
  {{Froese Fischer}}}, \bibinfo {author} {\bibfnamefont {S.}~\bibnamefont
  {Verdebout}}, \bibinfo {author} {\bibfnamefont {M.}~\bibnamefont
  {Godefroid}}, \bibinfo {author} {\bibfnamefont {P.}~\bibnamefont {Rynkun}},
  \bibinfo {author} {\bibfnamefont {P.}~\bibnamefont {J\"{o}nsson}}, \ and\
  \bibinfo {author} {\bibfnamefont {G.}~\bibnamefont {Gaigalas}},\ }\href
  {\doibase 10.1103/PhysRevA.88.062506} {\bibfield  {journal} {\bibinfo
  {journal} {Physical Review A}\ }\textbf {\bibinfo {volume} {88}},\ \bibinfo
  {pages} {062506} (\bibinfo {year} {2013})}\BibitemShut {NoStop}%
\bibitem [{\citenamefont {Verdebout}\ \emph {et~al.}(2013)\citenamefont
  {Verdebout}, \citenamefont {Rynkun}, \citenamefont {J\"{o}nsson},
  \citenamefont {Gaigalas}, \citenamefont {{Froese Fischer}},\ and\
  \citenamefont {Godefroid}}]{Verdebout2013}%
  \BibitemOpen
  \bibfield  {author} {\bibinfo {author} {\bibfnamefont {S.}~\bibnamefont
  {Verdebout}}, \bibinfo {author} {\bibfnamefont {P.}~\bibnamefont {Rynkun}},
  \bibinfo {author} {\bibfnamefont {P.}~\bibnamefont {J\"{o}nsson}}, \bibinfo
  {author} {\bibfnamefont {G.}~\bibnamefont {Gaigalas}}, \bibinfo {author}
  {\bibfnamefont {C.}~\bibnamefont {{Froese Fischer}}}, \ and\ \bibinfo
  {author} {\bibfnamefont {M.}~\bibnamefont {Godefroid}},\ }\href {\doibase
  10.1088/0953-4075/46/8/085003} {\bibfield  {journal} {\bibinfo  {journal}
  {Journal of Physics B}\ }\textbf {\bibinfo {volume} {46}},\ \bibinfo {pages}
  {085003} (\bibinfo {year} {2013})}\BibitemShut {NoStop}%
\end{thebibliography}%

\end{document}